\documentclass[
    ,final            % use final for the camera ready runs
  ]
  {aipproc}

\layoutstyle{6x9}

\usepackage{amsfonts,amsmath,amssymb}
\usepackage{mathrsfs}
\usepackage{amscd}
\usepackage{color}
\usepackage{graphicx} 
\usepackage{hhline}
\usepackage{booktabs} %for wider rules tables need
\begin{document}

\title{Comparison of metrics obtained with analytic perturbation theory and a numerical code}

\classification{04.25.Nx, 04.40.Dg}
\keywords      {Relativistic astrophysics, post-Minkowskian approximation, Harmonic coordinates, Rotating stars}

\author{J. E. Cuch\'i}{
  address={Dpto. F\'isica Fundamental. Universidad de Salamanca}
}

\author{A. Molina}{
  address={Dpto.\ de F\'\i sica Fonamental. Universitat de Barcelona}
}

\author{E. Ruiz}{
  address={Dpto. F\'isica Fundamental. Universidad de Salamanca}
  } % additional visiting address}

\begin{abstract}
We compare metrics obtained through analytic perturbation theory with their numerical counterparts. The analytic solutions are computed with the CMMR post-Minkowskian and slow rotation approximation due to \citet{cabezas2007ags} for an asymptotically flat stationary spacetime containing a rotating perfect fluid compact source. The same spacetime is studied with the AKM numerical multi-domain spectral code \citep{ansorg2002highly,ansorg2003highly} . We then study their differences inside the source, near the infinity and in the matching surface, or equivalently, the global character of the analytic perturbation scheme.
\end{abstract}

\maketitle

%%%%%%%%%%%%%%%%%%%%%%%%%%%%%%%%%%%%%%%%%%%%
%% MAINMATTER
%%%%%%%%%%%%%%%%%%%%%%%%%%%%%%%%%%%%%%%%%%%%

\section{Introduction}

Despite the great effort invested, there is still no exact solution of Einstein's equations able to describe a stellar model, i.e., a singularity-free rotating body that has been matched to an asymptotically flat vacuum exterior. In the last two decades,
% accompanied by the flourishing of numerical techniques and capabilities, 
the attention has moved to the field of approximate solutions.
% While analytic approximations are as old as General Relativity itself, the work of \citet{wilson1972models} was the first of many numerical schemes to follow. 
Among the recent ones is the AKM code \citep{ansorg2002highly,ansorg2003highly}. It is a multi-domain spectral method, and the difficulties many other codes have on surfaces of discontinuity of some sources due to Gibbs phenomena are solved computing the solution of the different domains and then imposing matching conditions. The number $n$ of Chebyshev polynomials in the expansions, resolution and equation of state (EOS) can be chosen, and reaches machine accuracy for high enough $n$. We will use it to check the behaviour of the CMMR post-Minkowskian+slow rotation analytic approximation scheme some of us introduced in \citep{cabezas2007ags}. We begin fixing the properties of the problem spacetime while briefing the CMMR basics. Then we study the relative error in the metric functions between schemes  and some physical properties of the source. 

\section{Building the metrics}

\begin{figure}
\begin{tabular}{cc}
 \includegraphics[width=210pt]{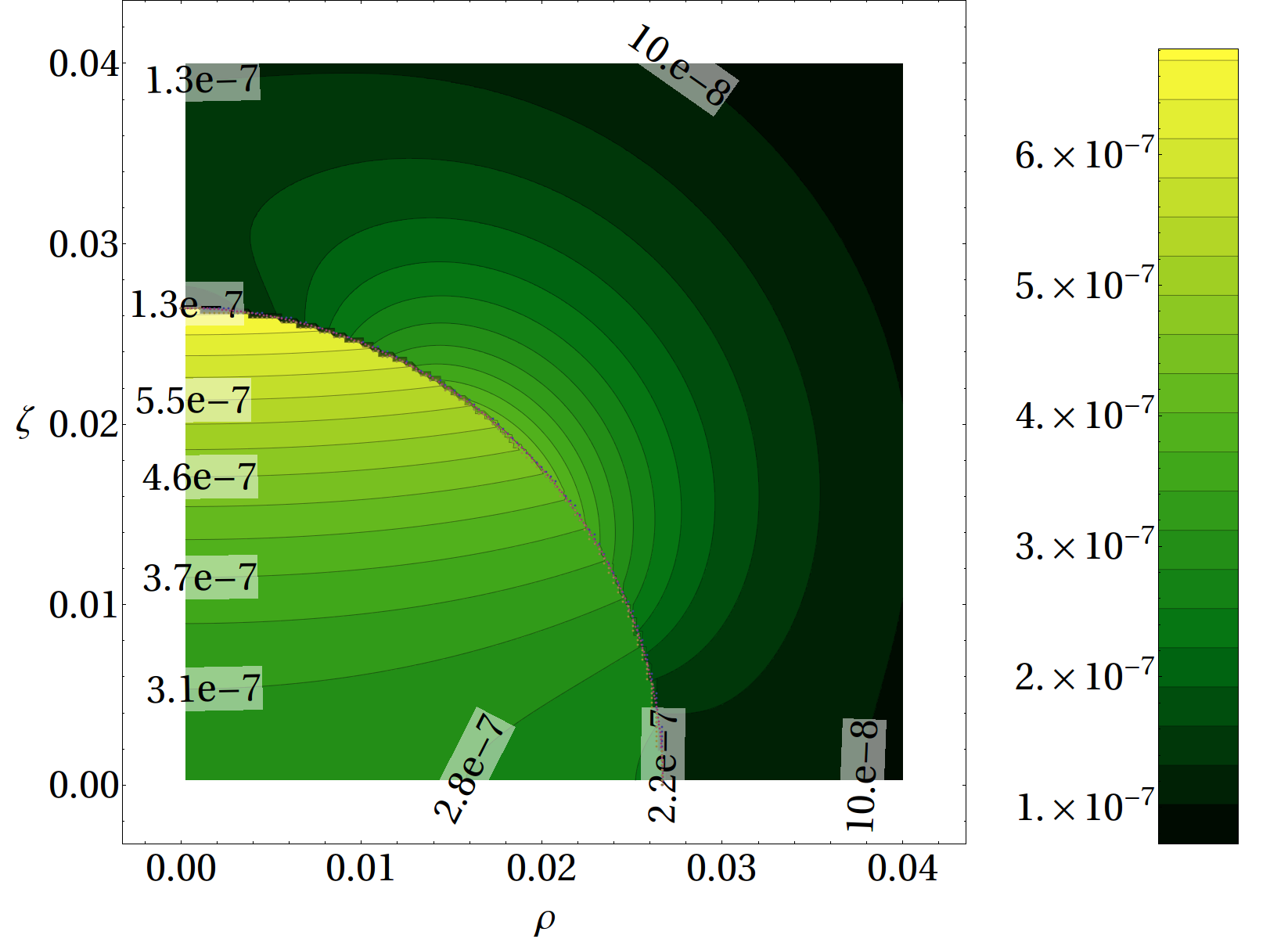}&
\includegraphics[width=208pt]{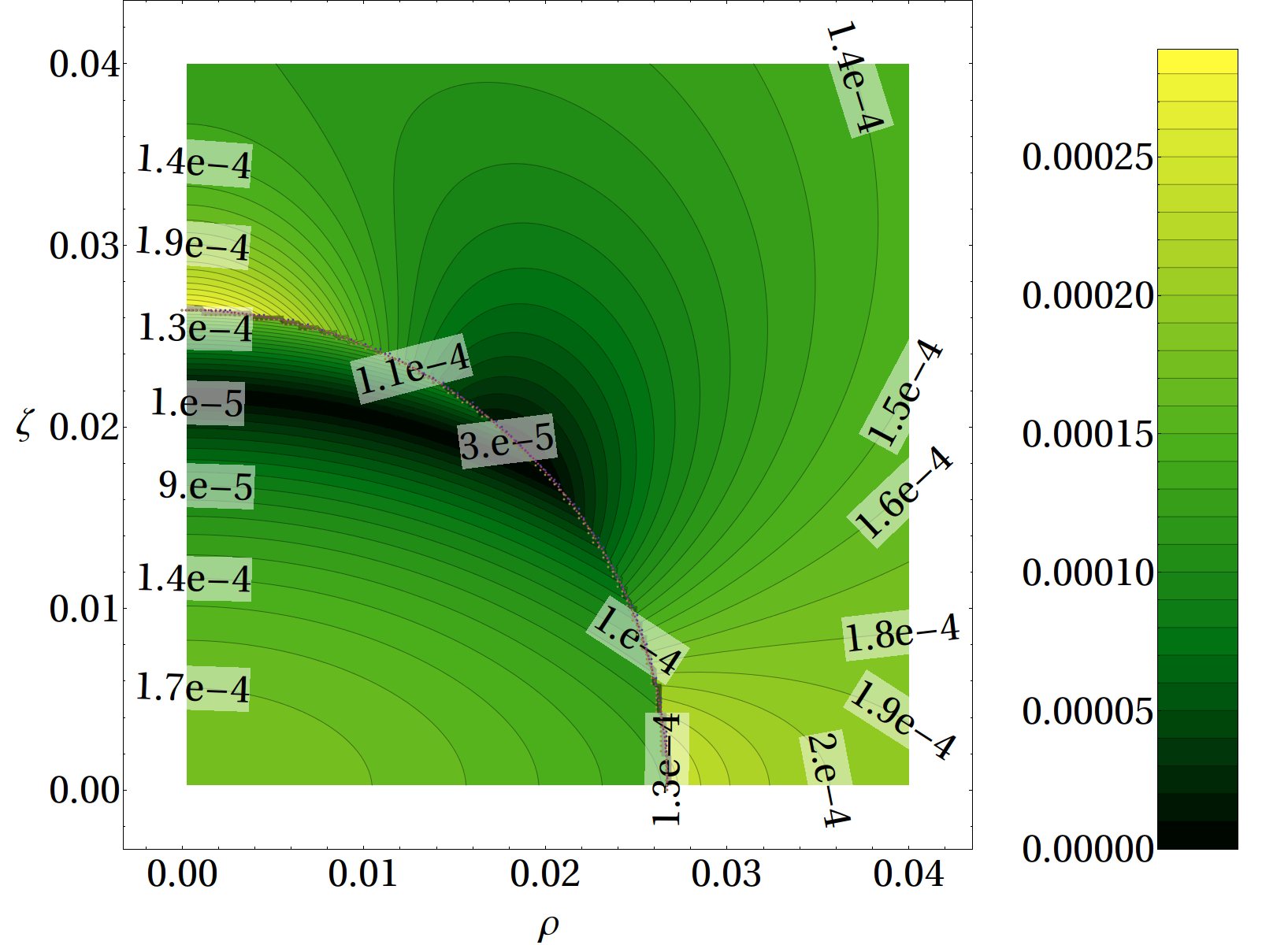}\\[-2ex]
\begin{small}(a)\end{small}&\begin{small}(b)\end{small}\\
\includegraphics[width=210pt]{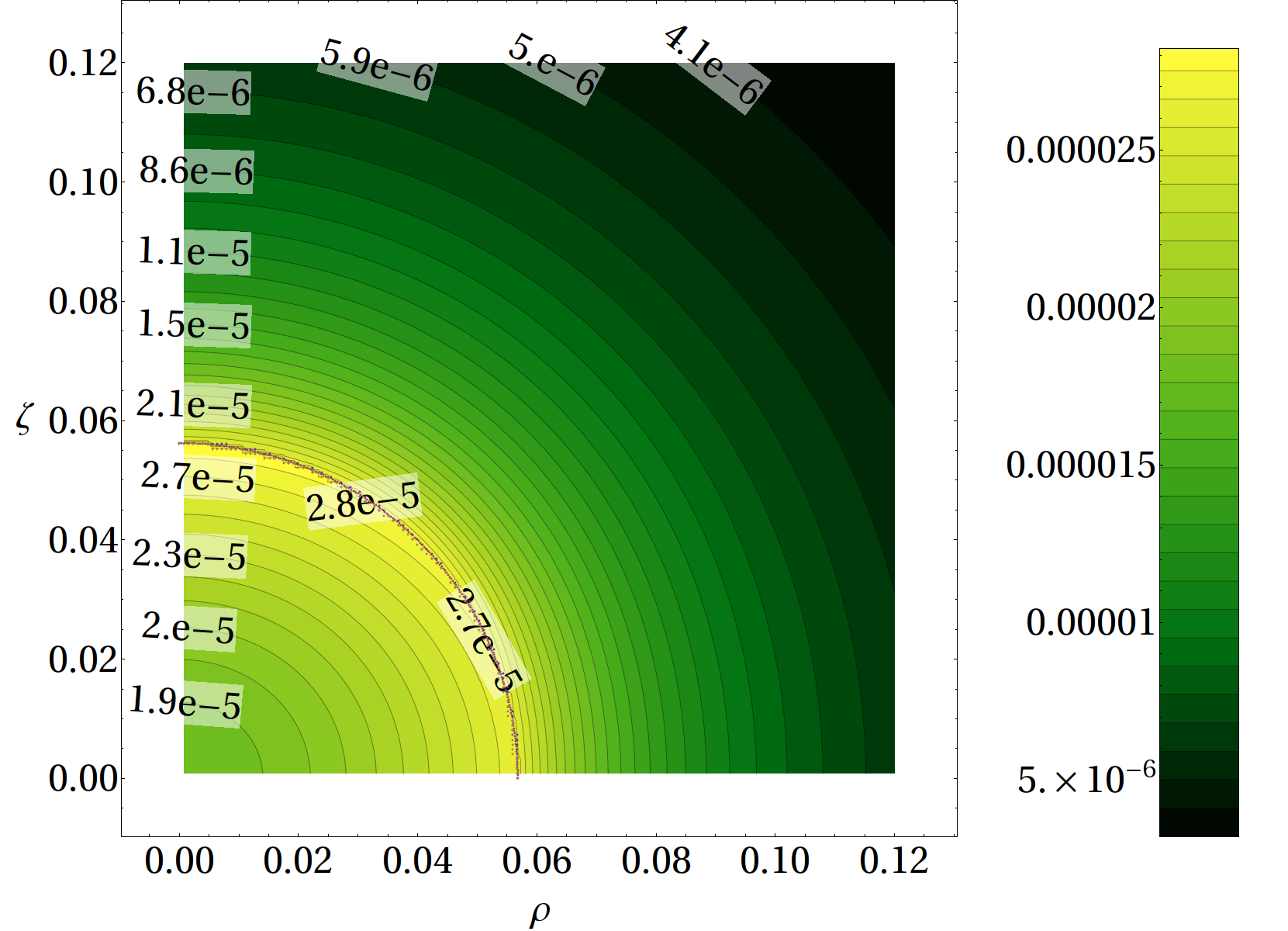}&
\includegraphics[width=203pt]{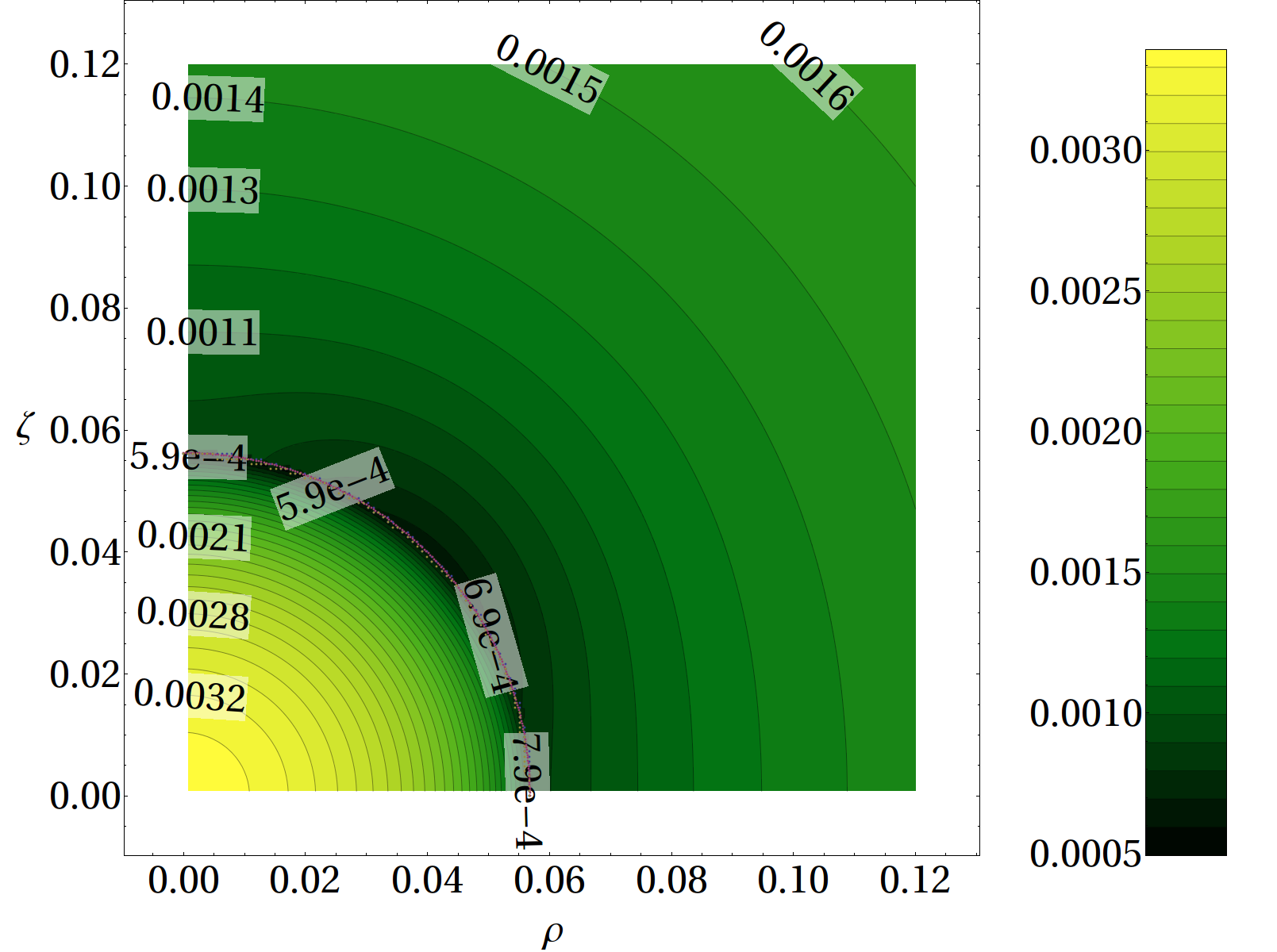}\\[-2ex]
\begin{small}(c)\end{small}&\begin{small}(d)\end{small}
\end{tabular}
\caption{Relative error between CMMR and AKM in (a-b) $g_{tt}$ and $g_{t\varphi}$, $M_B=8\times10^{-5}$, $\omega=0.2$ $(\lambda\approx0.0030,\,\Omega\approx0.098)$; (c-d) $g_{tt}$ and $g_{t\varphi}$, $\smash{M_B=8\times10^{-4}}$, $\omega=0.2$ $(\lambda\approx0.013,\,\Omega\approx0.098)$. Note that both the scale and the angular dependence of the error decrease with bigger $M_B$. The thin dotted lines represent the AKM and CMMR surfaces (indistinguishable in this picture size).}
\label{fig1}
\end{figure}

\begin{figure}[b]
\includegraphics[width=215pt]{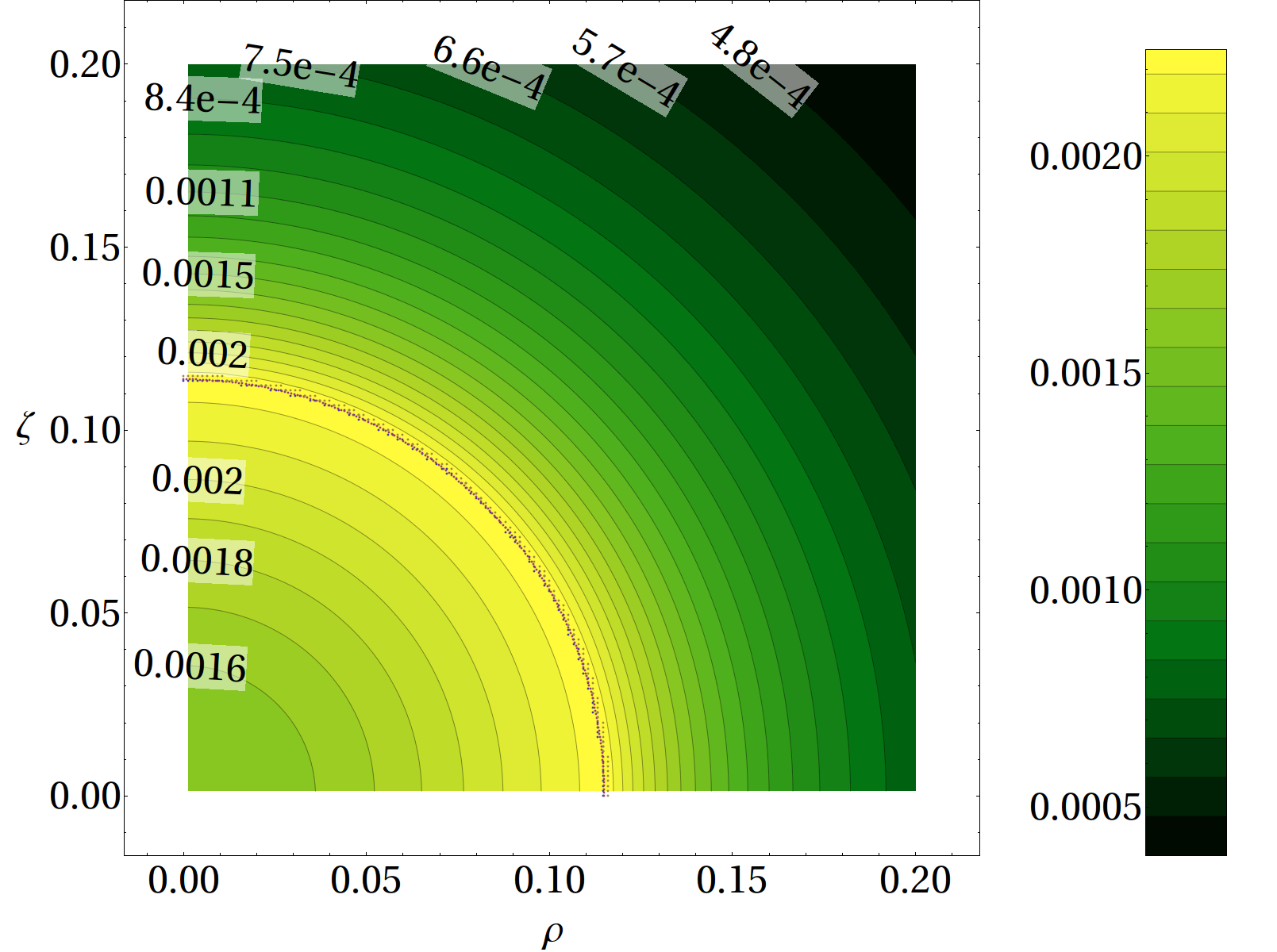}
\includegraphics[width=208pt]{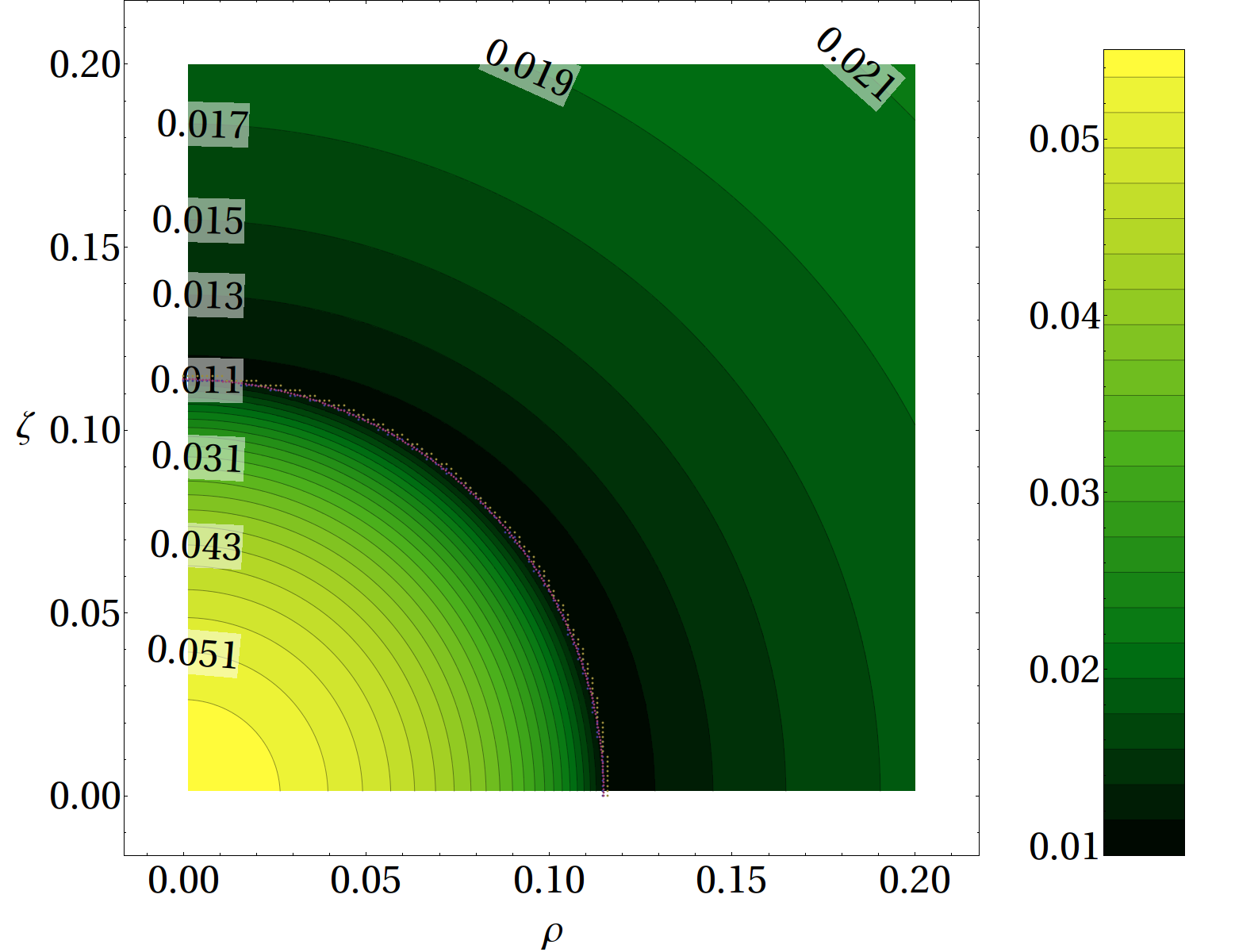}
\caption{Relative error between CMMR and AKM in $g_{tt}$ and $g_{t\varphi}$ for $M_B=8\times10^{-3}$, $\omega=0.2$ $(\lambda\approx0.056,\,\Omega\approx0.098)$. (Continued from Fig. 1)}
\label{fig2}
\end{figure}

The spacetime studied $\mathcal{V}$  is stationary, with timelike Killing vector field $\pmb{\xi}$, and axisymmetric, being $\pmb{\eta}$ the associated closed-orbits spacelike Killing vector field that satisfies regularity condition near the axis. It is built from the matching of two  spacetimes.
 The first one, $\mathcal{V}^-$, is filled with a perfect fluid in circular flow so that its velocity can be written $\mathbf{u}=\psi\left( \pmb{\xi}+\omega \pmb{\eta} \right)$, with $\psi$ adjusted to make $u^\alpha u_\alpha=-1$. The function $\omega$ is constant, making the fluid rigidly rotating. It has constant energy density, $\mu=\mu_0$, so integrating Euler's equations
%\citep{boyer1965rfm}
 gives the pressure $ p=\mu_0\left[ ({\psi}/{\psi_{\Sigma}}) -1\right]$, with $\psi=\psi_\Sigma$ for $p=0$.
% \begin{align}
%  p=\mu_0\left( \frac{\psi}{\psi_{\Sigma}} -1\right)&&(\psi_\Sigma=\psi\Leftrightarrow p=0).
% \end{align} 
The second spacetime, $\mathcal{V}^+$, is asymptotically flat vacuum surrounding $\mathcal{V}^-$.
It is not restrictive for us to use global harmonic Cartesian-like coordinates $\{x^\alpha\}$ in $\mathcal{V}$\citep{cuchi2009matching}. Working in spherical-like coordinates associated to $\{x^\alpha\}$, the $p=0$ surface $\Sigma$ on which $\mathcal{V}^-$ and $\mathcal{V}^+$ are matched can be written as an expansion $\smash{r_\Sigma=r_s \left[1+\sigma\; \Omega^2 P_2(\cos\theta)\right] +\mathcal{O}(\Omega^4)}$
% \begin{equation}
% r_\Sigma=r_s \left[1+S\; \Omega^2 P_2(\cos\theta)\right] +\mathcal{O}(\Omega^4)
% \label{sigma}\end{equation} 
in Legendre polynomials $P_n$, with $\sigma$ a constant. It has been truncated introducing a slow rotation approximation parameter $\Omega$ we have chosen as $\Omega^2=\omega^2 r_s^3/m$, where $m\equiv\frac43\pi \mu_0 r_s^3$ is the Newtonian mass of a sphere of radius $r_s$.

To solve Einstein's equations, we use a multipolar post-Minkowskian approximation as follows. Defining a parameter $\lambda=m/r_s$, the exact metric in each spacetime $\mathbf{g}^\pm$ is decomposed as $\mathbf{g}^\pm(\lambda,\Omega)=\pmb{\eta}+\,\mathbf{h}^\pm(\lambda,\Omega)$, with $\pmb{\eta}$ the flat metric.
%  and the expansion 
% \begin{equation}
% \mathbf{h}^\pm(\lambda,\Omega)=\sum_{n=0}^\infty\lambda^{1+\frac{n}{2}} \mathbf{h^\pm}^{(1+\frac{n}{2})}(\Omega), \quad n\in \mathbb{N} . \end{equation} 
Then, Einstein's equations are solved iteratively in $\lambda$. Both $\mathbf{h}^\pm(\lambda,\Omega)$ are tensor spherical harmonic expansions that are truncated, in this case, to contain $\Omega$ powers lower than $\Omega^4$. This restricts the number of multipole moments $M_i,\,J_{i+1}$ appearing in the exterior solution. We then match $\mathcal{V}^-$ and $\mathcal{V^+}$ imposing continuity of the metric and its first derivatives. %\citep{lichnerowicz1955theories} 
This fixes all coefficients and the stellar model depends then only on $\mu_0,$ $\omega$ and $r_s$.

The AKM code computes the matched value of the functions $U,\,k,\,W\text{ and }a$ in the general line element of a stationary axisymmetric perfect fluid or asymptotically flat vacuum spacetime in quasi-isotropic coordinates (see, e.g. \citep{ansorg2002highly}, where $\{\rho,\,\zeta \}$ are cylindrical associated to quasi-isotropic coordinates $\{r,\,\theta\}$)
% \begin{equation}
%  \dif{s}^2=\e^{-2U}\left[\e^{2k}\left( \dif{\rho}^2+\dif{\zeta}^2\right)+W^2\dif{\varphi}^2 \right]-\e^{2U}\left( \dif{t}+a\,\dif{\varphi} \right)^2
% \end{equation} 
at each point of a  coordinate grid of user-definable resolution. It also gives a lot of information in terms of physical and geometric parameters, such as multipolar moments $M_0$ and $J_1$, baryonic mass $M_B$, angular velocity $\omega$, equatorial radius $r_e$ and central pressure $p_c$ among others. Once the values of two of them and the EOS have been fixed, the code can compute the metric.

\section{Comparison results}
To compare the results of CMMR and AKM for $\mu=\mu_0$, we must first find the change of coordinates from the spherical-like ones of CMMR to quasi-isotropic. This change is necessarily approximate, introducing a new source of error in the comparison. This makes the relative error we compute between the metric functions of each scheme at a point to be a strict upper bound. 
For the comparison, CMMR was computed up to order $(\mathcal{O}(\lambda^{5/2}),\mathcal{O}(\Omega^3))$,  and AKM was set to use 12 Chebyshev polynomials in each direction.
Then, working in dimensionless quantities ($G=c=\mu_0=1$) we must choose which two parameters to fix in both CMMR and AKM. For this work, we have dealt with two sets, first $\{M_0,\,\omega\}$ and then $\{r_e,\,\omega\}$. Once $\omega$ is fixed, CMMR results depend only on $r_s$. We get its value equating both $M_0$ (alternatively, $r_e$) values. The $M_0$ adjustment gives better results and is the one we will focus on. Figs. \ref{fig1}-\ref{fig2} show the relative error in $g_{tt}$ and $g_{t\varphi}$ ($g_{ii}$ plots are very similar to $g_{tt}$ ones) on a quadrant of the plane $\rho-\zeta$ for  $\omega=0.2$ and AKM values of $M_B=8\times10^{-5}$, $8\times10^{-4}$ and $8\times10^{-3}$. For a typical neutron star density $\mu_0=4\times10^{17}\,\text{kg}\,\text{m}^{-3}$, they would correspond to a frequency $\nu\approx1033\,\text{s}^{-1}$ and $M_0\approx0.003M_\odot,\,0.03M_\odot$ and $0.3M_\odot$, respectively.
Table \ref{tab1} shows their CMMR values and relative errors of some quantities. The rather extreme cases of $\omega=0.7$ are included to check the behaviour of our slow rotation approximation for high values of $\Omega$ $(\Omega\approx0.49\omega)$.

\begin{table}
\begin{tabular}{lllllll}
\toprule
		&\text{CMMR}	&$\epsilon$	&\text{CMMR}	&$\epsilon$	&\text{CMMR}	&$\epsilon$ \\
\midrule
%$M_B$	&8.0\,e-5	&	&8\,e-5	&	&8\,e-4	&	\\
$\omega$	&0.2	&	&0.7	&	&0.2	&\\
$M_0$	&0.00007985	&	&0.00007986	&	&0.00079334	&	\\
$J_1$	&4.6065620\,e-9	&0.00020	&1.75490113\,e-8	&0.017	&2.13041342\,e-7	&0.0021	\\
$r_{e}$	&0.02674082	&4.4\,e-6	&0.02792957	&0.0074	&0.05689761	&0.00073\\
$p_c$	 &0.00147636	&0.012	&0.00136990	&0.00041	&0.00668441	&0.056	\\
$r_s$	&0.02663511	&	&0.02663467	&	&0.05667470	&	\\
\midrule 
%$M_B$	 &8\,e-4	&	&8\,e-3	&	&8\,e-3	&\\
$\omega$	 	&0.7	&	&0.2	&	&0.7	&\\
$M_0$	 &0.00079362	&	&0.00769003	&	&0.00770289	&\\
$J_1$	  &8.10712355\,e-7	&0.018	&9.52547199\,e-6	&0.032	&0.00003614	&0.043\\
$r_{e}$	  	&0.05940141	&0.0065	&0.11664607	&0.014	&0.12158066	&0.0080\\
$p_c$	 &0.00620181	&0.044	&0.02810237	&0.27	&0.02607038	&0.26\\
$r_s$	  &0.05667105	&	&0.11620612	&	&0.11619185	&\\
\bottomrule 
% 
% Recommended & yes & yes & no & no & no\\
% \hline
\end{tabular}
\caption{CMMR values of some quantities and relative error with AKM $\epsilon=\frac{|CMMR-AKM|}{AKM}$ for two members ($\omega=0.2$ and $\omega=0.7$) from each studied sequence: $M_B=8\times10^{-5}$, $8\times10^{-4}$ and $8\times10^{-3}$ }
\label{tab1}
\end{table}

%\section{Discussion}
 For the three cases studied, relative errors in metric functions increase roughly two orders of magnitude if we make $M_B$ ten times bigger, being higher for $g_{t\varphi}$ and the interiors. This is expectable since we have fixed $M_0$, i.e. the behaviour of $g_{tt}$ near spatial infinity, what can cause the high values of $\epsilon(p_c)$  as well. Significant error discontinuities are located on equatorial/polar lobes (Fig. 1b) and can be caused by the truncation at $\mathcal{O}(\Omega^3)$ of $\mathbf{h}^\pm$ and the Legendre expansion of $\Sigma$. This is supported by the smooth plots we get in the static limit with $r_e$ adjustment, and the increased lobular appearance when $M_B$ decreases (giving rise to more oblate configurations for the same $\omega$). We expect this angular dependence of $\epsilon$ to decrease including more terms of the $\Omega$ series.

The error inside the source is systematically bigger than outside it, but comparable. We plan to use $p_c$ and $J_1$ to fix $r_s$ and expect better results in the interior. We will add new physical quantities to the comparison to see how much the general performance improves going further in the approximation as well as other EOS.
%%%%%%%%%%%%%%%%%%%%%%%%%%%%%%%%%%%%%%%%%%%%%%%%
%% BACKMATTER
%%%%%%%%%%%%%%%%%%%%%%%%%%%%%%%%%%%%%%%%%%%%%%%%

\begin{theacknowledgments}
 We thank J. L. Jaramillo and J. Mart\'in for comments. JEC thanks the Albert Einstein Institut (Potsdam) for hospitality and Junta de Castilla y Le\'on for grant EDU/1165/2007.  This work was supported by grant FIS2009-07238 (MICINN).

\end{theacknowledgments}



\bibliographystyle{aipproc}   % if natbib is available

\bibliography{bibliograf}

% \IfFileExists{\jobname.bbl}{}
%  {\typeout{}
%   \typeout{******************************************}
%   \typeout{** Please run "bibtex \jobname" to optain}
%   \typeout{** the bibliography and then re-run LaTeX}
%   \typeout{** twice to fix the references!}
%   \typeout{******************************************}
%   \typeout{}
%  }

\end{document}